\documentstyle[12pt,epsfig]{article}

\begin{document}

\begin{flushright}
IMSc/2002/03/07 \\
hep-th/0204136
\end{flushright} 

\vspace{2ex}

\begin{center}

{\large \bf  }

\vspace{2ex}

{\large \bf  No Go Theorem for Self Tuning Solutions } \\

\vspace{2ex}

{\large \bf  With Gauss-Bonnet Terms} \\

\vspace{8ex}

{\large  Samik Dasgupta$^a$, Rajesh Venkatachalapathy$^b$, 
S. Kalyana Rama$^b$}

\vspace{3ex}

{\bf (a)} Department of Physics, Campus Box 390 University of Colorado, 

Boulder, CO 80309 USA.

{\bf (b)} Institute of Mathematical Sciences, C. I. T. Campus, 

Taramani, CHENNAI 600 113, India. 
 
\vspace{1ex}

email: samik.dasgupta@colorado.edu, \\ 
chinta@imsc.ernet.in, krama@imsc.ernet.in \\ 

\end{center}

\vspace{6ex}

\centerline{ABSTRACT}
\begin{quote}  
We consider self tuning solutions for a brane embedded in an
anti de Sitter spacetime. We include the higher derivative
Gauss-Bonnet terms in the action and study singularity free
solutions with finite effective Newton's constant. Using the
methods of Csaki et al, we prove that such solutions, when
exist, always require a fine tuning among the brane parameters.
We then present a new method of analysis in which the
qualitative features of the solutions can be seen easily without
obtaining the solutions explicitly. Also, the origin of the fine
tuning is transparent in this method. 
\end{quote}




\newpage

%
%

\vspace{4ex}

{\bf 1.}  
Randall and Sundrum had proposed a model a few years ago
\cite{rs} where a $3 + 1 - $dimensional brane is embedded in a
five dimensional anti de Sitter (AdS) spacetime. In this model,
the brane can be thought of as our universe, with various observable
particles assumed to be confined to the brane except
graviton which can propagte in the extra fifth dimension also,
which is non compact. They showed that for a particular value of
the brane tension, with the induced metric on the brane being
flat and thus preserving the Poincare invariance on the brane,
the zero mode of the graviton is confined to the brane. The
effective four dimensional Planck mass $M_4$ is then finite.

Soon after, the authors of \cite{sundrumkachru} constructed
another model where a scalar field $\phi$ , with a potential
$V(\phi)$, can also propagate in the extra dimension.  They
showed, in this model, that (the zero mode of) the graviton is
confined to the brane, preserving the Poincare invariance on the
brane, for any value of the brane tension. Taking the brane
tenison to be the brane cosmological constant, it therefore
follows that a flat Minkowski metric on the brane is possible,
preserving Poincare invariance, for any value of the
cosmological constant. This, then, could be a solution to the
celebrated `cosmological constant problem' \cite{weinberg}.

However, the price one has to pay for this attractive feature is
the presence of singularities in the extra dimension at a finite
proper distance from the brane. Analysing the solutions
explicitly for various choices of $V(\phi)$, it is found that
the singularities can be avoided, but then (the zero mode of)
the gravity will not be confined on the brane unless the brane
tension, equivalently the cosmological constant, is tuned to a
specific value. See \cite{others} for other attempts to solve 
this problem. 

In this context, Csaki et al \cite{csaki} have proved a no go
theorem that no singularity free solution with finite $M_4$,
which ensures that gravity is confined on the brane, is possible
without a fine tuning. They considered an action for graviton
and a scalar field with terms containing atmost two derivatives.
Thus, their analysis leaves open a possibility that such
solutions may exist for more general action containg higher
derivative terms. Also, such terms are expected to appear
generically in the effective action upon including the quantum
effects of gravity.

Following such a line of reasoning, Low and Zee \cite{lowzee}
have considered action with higher derivative terms. They
considered a specific combination of such terms, namely the
Gauss Bonnet terms, for graviton because of its special features
well known in the literature \cite{zumino}. The other more recent 
aspects of such terms in the action within the context of brane world 
phenomenology have been analysed in \cite{gbothers} \cite{neup}. The 
authors \cite{lowzee} analysed various cases explicitly and found 
in all these cases that no singularity free solution with finite 
$M_4$ is possible without a fine tuning.

In this paper, we consider an action containg the higher
derivative terms for graviton in the specific Gauss Bonnet
combination. The action contains a bulk scalar field $\phi$
with a potential $V(\phi)$. The brane tension is taken to be an
arbitrary function of $\phi$. We then study whether a
singularity free solution with finite $M_4$ is possible without
a fine tuning. 

We prove, following closely the method of \cite{csaki}, 
that such a solution is not possible. This is the analogue 
of the no go theorem of \cite{csaki}, but now valid for the 
case where the action contains the higher derivative 
Gauss Bonnet terms. 

We then  present a new method of analysis of the equations
involved, reminiscent of the method of `phase space analysis'.
In this method, the qualitative features of the solutions, such
as the presence or absence of singularities, finiteness or
otherwise of $M_4$, etc, can be seen easily without obtaining
the solutions explicitly. This method is applicable quite
generally with or without the higher derivative Gauss Bonnet
terms. Moreover, it provides a constructive way of obtaining
potentials $V(\phi)$ which will admit singularity free solutions
with finite $M_4$. However, they will all require a fine tuning
whose origin is transparent in this method.

The plan of the paper is as follows. In section {\bf 2}, we set
up our notations, present the action, the equations of motion,
and the boundary conditions that the solutions must satisfy. In
section {\bf 3}, we establish the no go theorem following the
method of \cite{csaki}. In section {\bf 4}, we present the
method of `phase space analysis', concluding in section {\bf 5}
with a brief summary and a few remarks.

%
%

\vspace{4ex}

{\bf 2.}  
We consider $d = 4 + 1$ dimensional spacetime, with coordinates
$x^M = (x^\mu, y)$, where $\mu = 0, 1, 2, 3$, $- \infty \le x^M
\le \infty$, and with a $(3 + 1)$ dimensional brane located at
$y = 0$. The bulk fields are the graviton $G_{M N}$ and a scalar
field $\phi$. Their action, including a higher derivative
Gauss-Bonnet term for $G_{M N}$ and a potential $V$ for $\phi$,
is given by 
\begin{eqnarray} 
S & = & 
\frac{1}{2} \int d^5x \sqrt{G} \; (R \; 
+ 4 \; \lambda \; (R^2 - 4 R^{MN} R_{MN} + R^{MNPQ} R_{MNPQ}) 
\nonumber \\
& - & \frac{3}{4} (\partial_M \phi \partial^M \phi + V(\phi) 
+ f(\phi) \; \delta(y) \; ) \; ) \; ,  \label{s}  
\end{eqnarray} 
where the five dimensional Planck mass is set equal to unity,
and the last term denotes the effective tension of the brane
located at $y = 0$, with $f(\phi)$ an arbitrary function of
$\phi$. 

We consider warped spacetime solutions, preserving the Poincare
invariance along the brane directions $x^\mu$. Thus, the fields are
functions of $y$ only. The metric $G_{M N}$ can then be written as
\begin{equation}\label{ds}
ds^2 = e^{- \frac{A(y)}{2}} 
\eta_{\mu \nu} dx^{\mu} dx^{\nu} + dy^2  
\end{equation}
where $\eta_{\mu \nu} = diag (- 1, 1, 1, 1)$ and $A(y)$ is the
warp factor. The curvature invariants are all then functions of
$A(y)$ and its derivatives.  For example, the Ricci scalar 
$R = 2 W' - \frac{5 W^2}{4}$ where we have defined 
\begin{equation}\label{wa'}
W \equiv A' \; , 
\end{equation}
and the primes, here and in the following, denote 
differentiation with respect to $y$. Also, we set $A(0) = 0$
with no loss of generality. $A(y)$ is then given by 
\begin{equation}\label{a}
A(y) = \int_0^y dy \; W(y) \; . 
\end{equation}

The equations of motion, for $y \ne 0$, that follow 
from the action (\ref{s}) are 
\begin{eqnarray}
W' \; (1 - \lambda W^2) & = & \phi'^2 \label{w'} \\
W^2 \; (1 - \frac{\lambda W^2}{2}) & = &  \phi'^2 - V 
\label{wphi} \\  
2 \; (\phi'' - W \phi') & = & V_{(1)} \; , \label{phi''} 
\end{eqnarray} 
where the subscript $(n)$ denotes the $n^{th}$ differential with
respect to $\phi$.  \footnote{Generically, only two of the above
equations are indpendent: equations (\ref{wphi}) and (\ref{w'})
imply equation (\ref{phi''}) if $\phi' \ne 0$; equations
(\ref{wphi}) and (\ref{phi''}) imply equation (\ref{w'}) if 
$W \ne 0$. Equation (\ref{wphi}) is, if $\phi' \ne 0 $the `energy'
integral of motion for equations (\ref{w'}) and (\ref{phi''}) with the
integration constant set to zero.}

The presence of brane at $y = 0$ imposes the following boundary
conditions at $y = 0$: 
\begin{equation}\label{bc} 
\phi_+ = \phi_- \equiv \phi_0 
\; , \; \; \; \;  
\tilde{W}_+ - \tilde{W}_- = 2 a 
\; , \; \; \; \; 
\phi'_+ - \phi'_- = 2 b \; , 
\end{equation} 
where the subscripts $\pm$ denote the values at $y = 0_\pm$,
$\tilde{W} = (1 - \frac{\lambda W^2}{3}) \; W$, $2 a =
f(\phi_0)$, and $2 b = f_{(1)}(\phi_0)$. The parameters $a$ and
$b$ are arbitrary and independent constants since the function
$f(\phi)$ is arbitrary. If the values of $a$ and $b$ are
restricted, or constrained to obey any specific relation, then
they are said to be fine tuned.

Upon substituting the solution for $A$ in the bulk
action, and performing the $y$-integration, one can define an
effective four dimensional Planck mass $M_4$ \cite{lowzee} as follows: 
\begin{equation}\label{m4}
M_4^2 \propto \int_{y_L}^{y_R} dy \; e^{-A(y)/2} \; , 
\end{equation}
where $y_L < 0$ and $y_R > 0$ are the locations of the
singularity, if any, closest to the brane on either side. The exact 
expression which involves a $\lambda$ dependent factor is given in 
\cite{neup}. If there are no singularities then $y_L = - \infty$ and 
$y_R = + \infty$. It then follows \cite{rs} that if $M_4$ is
finite then (the zero mode of) the graviton is confined to the
four dimensional brane at $y = 0$.

Our main interest is to study the solutions to equations
(\ref{w'}) - (\ref{phi''}), satisfying the boundary conditions
(\ref{bc}). And, more specifically, to study whether the
solutions are free of singularities and have finite $M_4$.
Typically, solutions will have either infinite $M_4$, or
singularities at finite proper distance in the $y$ direction,
or both, depending on the potential $V(\phi)$ and the
values of the arbitrary constants $a$ and $b$. Singularity free
solutions with finite $M_4$ are possible, if at all, for only a
restricted class of the potentials $V(\phi)$. 

In \cite{csaki}, Csaki et al have proved, for actions containing
terms with atmost two derivatives, that a singularity free
solution with finite $M_4$, when exists, will always involve a
fine tuning.  However, effective actions incorporating quantum
gravity corrections typically contain terms with higher
derivatives. Then, perhaps, it may be possible to obtain
singularity free solutions with finite $M_4$, without any fine
tuning.

In this paper, we address this issue. Concretely, we include
higher derivative terms for the graviton in the form of
Gauss-Bonnet combination. The action is then given by equation
(\ref{s}). By an analysis similar to that of \cite{csaki}, we
look for solutions with finite $M_4$ and with no singularities,
and study whether any fine tuning is required.

We also present a method of analysis of the equations of
motion, reminiscent of the method of 'phase space analysis',
 where one can determine the presence or absence of
singularities, and the finiteness or otherwise of $M_4$, without
having to obtain explicitly the complete solution.  This method
can also be used to construct $V(\phi)$ which will admit
singularity free solution with finite $M_4$. It turns out that
such solution always involves one fine tuning, whose origin can
be seen clearly in this method.

%
%

\vspace{4ex}

{\bf 3.}  
To solve equations (\ref{w'}) - (\ref{phi''}), first 
consider $W$ as a function of $\phi$. Equations (\ref{w'}) 
and (\ref{wphi}) can then be written as 
\begin{eqnarray}
\phi' & = & (1 - \lambda W^2) \; W_{(1)} \label{phi'w} \\
V & = & (W_{(1)}^2 + \frac{1}{2 \lambda}) \; 
(1 - \lambda W^2)^2 - \frac{1}{2 \lambda} \; . 
\label{vw} 
\end{eqnarray} 
In principle, given the potential $V(\phi)$, equation (\ref{vw})
can be solved for $W(\phi)$; equation (\ref{phi'w}) can then be
solved for $\phi(y)$, and thus for $W(y)$; $A(y)$ is then given
by (\ref{a}) \cite{gubser}. 

In practice, however, this procedure is of limited use in
obtaining general solutions for a given $V(\phi)$ since equation
(\ref{vw}) is non linear.  Instead, one starts with a $W(\phi)$.
The corresponding potential $V(\phi)$ is given by equation
(\ref{vw}). Using equation (\ref{phi'w}), one then obtains
$\phi(y)$, and thus $W(y)$ and $A(y)$. Note that the solution
thus obtained is not the most general solution to equations
(\ref{w'}) - (\ref{phi''}) for the given $V(\phi)$.

We now study, following the method of \cite{csaki}, whether
equations (\ref{w'}) - (\ref{phi''}), equivalently (\ref{phi'w})
and (\ref{vw}), admit any singularity free solution with finite
$M_4$, for arbitrary values of the constants $a$ and $b$
with no constraints imposed on them - that is, with no fine
tuning.

We first define a conformal coordinate $z(y)$, with 
$z(0) = 0$, by 
\begin{equation}\label{z} 
y = \int_0^z d z \; \Omega(z) 
\; \; \; \; {\leftrightarrow} \; \; \; \; 
d y = \Omega d z  
\; \; \; \; {\rm where} \; \; \; \; 
\Omega(z) = e^{- \frac{A}{4}} \; . 
\end{equation} 
Then, $M_4$ and the proper distance, $l(y)$, from the brane
along the $y$-direction are given by 
\begin{equation}\label{l}
M_4^2 = \int_{z_L}^{z_R} dz \; \Omega^3 
\; \; \; \; 
{\rm and} \; \; \; \; 
l(y) = \int_0^y d y = \int_0^z d z \Omega(z) \; . 
\end{equation}
We assume that $M_4$ is finite and that there is no
singularity. Then $0 \le \vert y \vert \le \infty$ and, hence,  
$l(\infty) \to \infty$. Consider the limit $y \to \infty$. (The
limit $y \to - \infty$ can be analysed similarly.) Let
\[
\Omega \simeq K^{- 1} z^q 
\]
where $K$ is a positive constant. Then, in this limit, it
follows that 
\begin{eqnarray}
y \simeq \frac{z^{q + 1}}{K (q + 1)}  
\; , \; \; \; \; 
A(z) & \simeq & - \frac{4 q}{q + 1} \; ln \vert y \vert 
\; \; \; \; {\rm for} \; \; \; \; 
q \ne - 1   \label {aq} \\ 
y \simeq K^{- 1}ln z \; , \; \; \; \; 
A(z) & \simeq & 4 \; K \; \vert y \vert 
\; \; \; \; \; \; 
\; \; \; \;    
\; \; \;  {\rm for} \; \; \; \; 
q = - 1  \; .  \label{aq-1} 
\end{eqnarray}

The range of $q$ is severely restricted. The requirement that
$l(\infty) \to \infty$ and $M_4$ be finite implies that
\begin{equation}\label{qlimits}
- 1 \le q < - \frac{1}{3} \; . 
\end{equation}
Moreover, $W' (1 - \lambda W^2) \ge 0$ for all values of $y$
(see equation (\ref{w'})). For $q = - 1$, this inequality is
satisfied since $W' = 0$. For $q \ne - 1$, it implies that
\[
\frac{4 q}{K (q + 1) y^2} \; \left( 
1 - \frac{16 \lambda q^2}{K^2 (q + 1)^2 y^2} \right) \ge 0 \; . 
\] 

In the limit $\vert y \vert \to \infty$ that is being considered
here, the second factor is positive and, hence, $q$ must satsify
either $q \ge 0$ or $q < - 1$. Together, these constraints imply
that, in the limit $\vert y \vert \to \infty$, $q = - 1$ and
\begin{equation}\label{asymp}
A(y) \to 4 K \vert y \vert  \; , \; \; \; \; 
\phi' \to 0 \; , \; \; \; \; 
\phi \to \phi_c \; , 
\end{equation}
where $\phi_c \equiv \phi_R$ ($\phi_L$), for $y \to \infty$ 
($- \infty$), is a constant.  Using equations (\ref{wa'}), 
(\ref{phi'w}), (\ref{vw}), and (\ref{asymp}), it follows that,
at $\phi = \phi_c$, $W_{(1)} = 0$ and
\begin{equation}\label{sgn}
Sgn (W) = \sigma  \; , \; \; \; 
Sgn \left(W_{(n)} \; (1 - \lambda W^2) \; (\phi')^n
\right) = (- \sigma)^{n - 1} 
\end{equation}   
where $\sigma = Sgn(y)$ and $W_{(n)}$, $n > 1$, is the first
nonvanishing derivative of $W$ at $\phi_c$. The sign of $\phi'$,
required only when $n$ is odd, is obtained by evaluating $\phi'$
slightly away from $\phi_c$. The relations involving $W_{(1)}$
and $W$ follow directly from equations (\ref{phi'w}) and
(\ref{asymp}). To obtain that involving $W_{(n)}$, we considered
a few examples with different, but generic, $y$-dependences for
$W$ which satisfy equation (\ref{asymp}). $\phi'(y)$ and, thus
$\frac{d^n \phi}{d y^n}$, can then be obtained using equation
(\ref{w'}) and repeated differentiation. Now, equation
(\ref{phi'w}) gives 
\[
\frac{d^n \phi}{d y^n} = 
W_{(n)} \; (1 - \lambda W^2) \; (\phi')^{n - 1} 
\]
where $W_{(n)}$ is the first nonvanishing derivative of $W$ at
$\phi_c$.  Equation (\ref{sgn}) is obtained, in each of the
examples considered, upon comparing the two expressions for
$\frac{d^n \phi}{d y^n}$ thus derived.

We can now evaluate $W_{(n)}(\phi_c)$ for all $n$, in terms of
$V_{(m)}(\phi_c) = \frac{d^m V}{d \phi^m}(\phi_c)$, using
equation (\ref{vw}). $W(\phi)$ is then given by 
\begin{equation}\label{w}
W(\phi) = \sum_{k = 0}^\infty 
\frac{W_{(k)}}{k !} \; (\phi - \phi_c)^k 
\end{equation} 
which, by construction, solves the equation (\ref{vw}) in the
region $0 \le y \le \infty$ for $\phi_c = \phi_R$, and in the
region $- \infty \le y \le 0$ for $\phi_c = \phi_L$. (Here and
in the following, the coefficients in the Taylor expansions are
all to be evaluated at $\phi_c$. Their argument $\phi_c$
will not be written explicitly.) Equations (\ref{wa'}) and
(\ref{phi'w}) can then be used to obtain, in principle, the
complete solutions $\phi(y)$ and $A(y)$ for $0 \le \vert y \vert
\le \infty$. The resulting solution is, by construction,
singularity free with finite $M_4$.

Now, $V_{(n)}$ can be written as 
\begin{equation}\label{vfg} 
V_{(n)} = \sum_{k = 0}^n \frac{n !}{k ! (n - k) !} \; 
f_{(k)} \; g_{(n - k)} 
\end{equation}
where 
\[
f_{(0)} = f \equiv W_{(1)}^2 + \frac{1}{2 \lambda} 
\; , \; \; \; \; \;  
g_{(0)} = g \equiv (1 - \lambda W^2)^2 \; . 
\]
For $n = 0$, $1$, and $2$, we get, at $\phi = \phi_c$, 
\begin{eqnarray*}
V & = & - \; W^2 \; (1 - \frac{\lambda W^2}{2}) 
\; , \;  \; \; \;  \;  \; \; \; \; V_{(1)} = 0 \; , \\ 
V_{(2)} & = & 2 \; W_{(2)} \; (1 - \lambda W^2) \; \left( 
W_{(2)} \; (1 - \lambda W^2) - W \right) 
\end{eqnarray*}
where we have used $W_{(1)} = 0$.  $W_{(2)}$ can be solved in
terms of $V_{(2)}$ and admits two branches. However, equations
(\ref{sgn}) rule out one branch and imply, furthermore, that
\begin{eqnarray} 
V_{(2)} (\phi_c) & \ge & 0 \nonumber \\
2 W_{(2)} (1 - \lambda W^2) & = & W - \sqrt{W^2 + 2 V_{(2)}} 
\; \; \; {\rm for} \; \; \; \phi_c = \phi_R  \nonumber \\
2 W_{(2)} (1 - \lambda W^2) & = & W + \sqrt{W^2 + 2 V_{(2)}} 
\; \; \; {\rm for} \; \; \; \phi_c = \phi_L  \; . 
\label{v2}
\end{eqnarray} 
For $n > 2$, similar expressions can be obtained relating
$W_{(n)}$ and $V_{(n)}$. The highest derivative of $W$ that
will appear on the right hand side of equation (\ref{vfg}) is
$W_{(n)}$ since $W_{(1)} = 0$. Its coefficient can be easily
seen to be given by
\begin{equation}\label{wn}
V_{(n)} = 2 \; W_{(n)} \; (1 - \lambda W^2) \; \left( 
n \; W_{(2)} \; (1 - \lambda W^2) - W \right) + \cdots 
\end{equation} 
where $\cdots$ represent terms involving $W_{(k)}$ with $k < n$,
which can all be obtained explicitly by a straightforward
combinatorics, but are not necessary for our purposes here.

Note that the coefficient of $W_{(n)}$ never vanishes
\footnote{except when $(1 - \lambda W^2) = 0$, a special case
which can be easily analysed, with no change in the
conclusions.}. This is because $n$ is positive, and $W_{(2)} (1
- \lambda W^2) \le 0$ ($ \ge 0$) when $W > 0$ ($ < 0$), which
follows from equation (\ref{sgn}). Therefore, equation
(\ref{wn}) can be inverted to obtain $W_{(n)}$ in terms of
$V_{(n)}$ and $W_{(k)}$'s, with $k < n$. These relations can be
used iteratively to express $W_{(n)}$ in terms of $V_{(k)}$, $k
\le n$, for all $n$. Equation (\ref{w}) then gives the functions
$W_R(\phi)$ and $W_L(\phi)$ which, by construction, solves the
equation (\ref{vw}) in the region $0 \le y \le \infty$ for
$\phi_c = \phi_R$, and in the region $- \infty \le y \le 0$ for
$\phi_c = \phi_L$ respectively. Thus 
\begin{eqnarray}
W_R(\phi) & = & \sum_{k = 0}^\infty 
\frac{W_{(k)}(\phi_R)}{k !} \; (\phi - \phi_R)^k \nonumber \\
W_L(\phi) & = & \sum_{k = 0}^\infty 
\frac{W_{(k)}(\phi_L)}{k !} \; (\phi - \phi_L)^k \; . 
\label{wrl}
\end{eqnarray} 

The boundary conditions (\ref{bc}) are yet to be imposed where 
the parameters $a$ and $b$ are arbitrary constants. The equation 
involving $a$ implies that 
\[
\left(1 - \frac{\lambda W_R^2(\phi_0)}{3}\right) W_R(\phi_0) 
- \left(1 - \frac{\lambda W_L^2(\phi_0)}{3}\right)
W_L(\phi_0) = 2 a 
\]
where $W_R(\phi)$ and $W_L(\phi)$ are given by equations
(\ref{wrl}). This condition fixes the value of $\phi_0$ in terms
of the parameter $a$. Once $\phi_0$ is fixed, the discontinuity
in $\phi'$, namely, 
\[
\phi'_+ - \phi'_- = 
\left(1 - \lambda W_R^2(\phi_0)\right) W_{R(1)}(\phi_0) 
- \left(1 - \lambda W_L^2(\phi_0)\right) W_{L(1)}(\phi_0) 
\]
is also fixed. This, in turn, implies that the parameter $b$ can
not be arbitrary, but must be related to the parameter $a$ as
given by the above set of equations.  Hence, it follows that
singularity free solution(s) with finite $M_4$ will involve a
fine tuning.

Note that $\phi_R$ and $\phi_L$ are not arbitrary but must be
the values of $\phi$ for which $V(\phi)$ is a minimum.  (See
equation ({\ref{v2}).) Thus, they can take only a discrete set
of values and, in particular, cannot be continuos parameters of
the solution. The boundary conditions above may also allow for a
discrete set of values for $\phi_0$. Hence, a similar discrete
range of values is also allowed for the parameter $b$.
Nevertheless, the values of $a$ and $b$ must be related as given
by the above set of equations and, hence, there must be a fine
tuning. Various subtleties that may arise regarding the choice
of the branch(es) of $W(\phi)$, etc. are analysed in detail in
\cite{csaki}. The same analysis remains valid in the present
case also, where the higher derivative Gauss-Bonnet term is
included in the action. Thus, we conclude that singularity free
solutions with finite $M_4$ requires a fine tuning. This is the analogue
of the no go theorem of [5], but now valid for the case when the action
contains higher derivative Gauss Bonnet terms.

%
%

\vspace{4ex}

{\bf 4.}
We now present another method of analysis of the equations
involved, reminiscent of the method of `phase space analysis'.
In this method, the qualitative features of the solutions, such
as the presence or absence of singularities, finiteness or
otherwise of $M_4$, etc, can be seen easily without obtaining
the solutions explicitly. This method also provides a
constructive way of obtaining potentials $V(\phi)$ which will
admit singularity free solutions with finite $M_4$. However,
they will all require a fine tuning whose origin is transparent
in this method.
  
Consider $W$ as a function of $\phi$. Then one can, in
principle, obtain $\phi(W)$ and thus $W_{(1)}(W)$ as functions
of $W$.  \footnote{For example, if $W = \alpha + \beta \phi^n$
then $W_{(1)} = n \beta \left( \frac{W - \alpha}{\beta}
\right)^{\frac{n - 1}{n}}$.}  $V(W)$ is then given by
(\ref{vw}). It follows, from equations (\ref{w'}) and
(\ref{wphi}), that 
\begin{equation}\label{w'v}  
W' = \frac{V + W^2 \; (1 - \frac{\lambda W^2}{2})}
{1 - \lambda W^2} 
\; \; \; \; {\rm with} \; \; \; 
V + W^2 \; (1 - \frac{\lambda W^2}{2}) \ge 0 \; ,  
\end{equation} 
from which $W(y)$ can, in principle, be obtained since $V$ is
now a function of $W$. $\phi(y)$ follows since $\phi$ too is a
function of $W$.

Using $d y = d W \; \left(\frac{1}{W'}\right)$, 
with $W'(W)$ given by (\ref{w'v}), we have  
\begin{eqnarray*}
& & y = \int d W \; \left(\frac{1}{W'}\right)
\; \; , \; \; \; \; \; 
A = \int d W \; \left(\frac{W}{W'}\right) 
\; \; , \; \; \; \; \; 
M_4^2 = \int d W \; \left(
\frac{e^{- \frac{A}{2}}}{W'}\right) \\ 
& & \phi = \int d W \; \frac{1}{W'} \; 
\left( V + W^2 \; (1 - \frac{\lambda W^2}{2}) 
\right)^{\frac{1}{2}} \; . 
\end{eqnarray*}
Thus, the required quantities are all written as integrals of
functions of $W$ with respect to $W$. Then, various qualitative
features, such as presence or absence of singularities,
finiteness or otherwise of $M_4$ can all be obtained,
essentially, by inspection and simple asymptotice analyses in
suitable limits.

We illustrate this approach by an example. Let $V = - \alpha^2$,
with $0 < 2 \lambda \alpha^2 < 1$. The plot of $\frac{d y}{d W}
= \frac{1}{W'}$ vs $W$ is given in Figure 1 where only the
regions satisfying $W^2 \; (1 - \frac{\lambda W^2}{2}) \ge
\alpha^2$ are allowed. That is, the allowed regions are given by 
\[
\vert a \vert \le \vert W \vert 
\le \vert b \vert 
\]
where $a$ and $b$, both taken to be positive, are given by 
\[
a^2 = \frac{1 - \sqrt{1 - 2 \lambda \alpha^2}}{\lambda} 
\; , \; \; \; \; 
b^2 = \frac{1 + \sqrt{1 - 2 \lambda \alpha^2}}{\lambda} \; . 
\]
\begin{figure}[htb]
\begin{center}
\epsfig{file=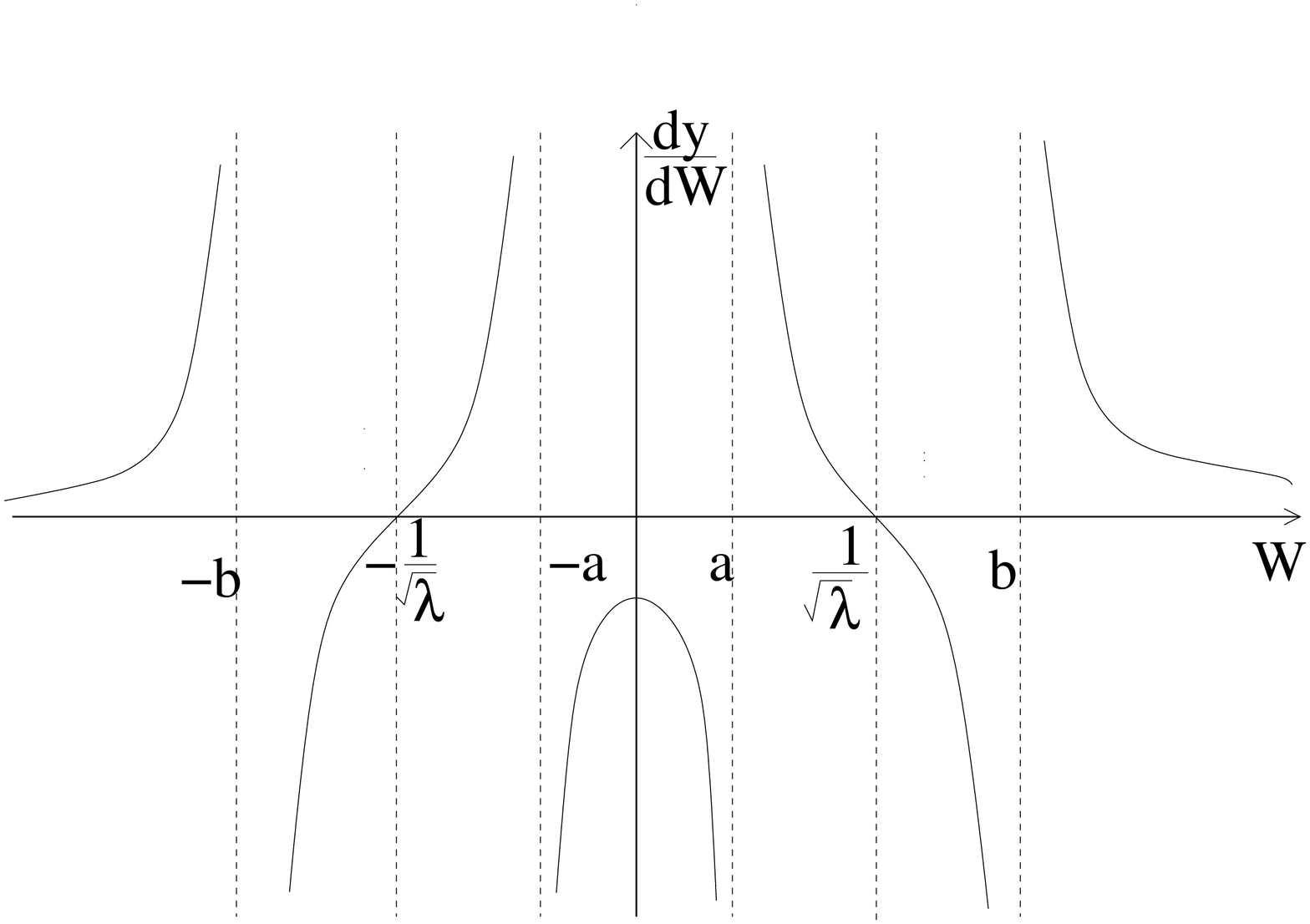, width=12cm, angle=0}
\end{center}
\caption{Plot of $d y / d W$ vs $W$, for $V = -\alpha^2$. Allowed
regions are $|a| \le |W| \le |b|$.}
\end{figure}

The plot of $y$ vs $W$ can then be obtained from Figure 1 by
inspection and a simple asymptotic analysis in the limits
$\vert\frac{d y}{d W}\vert \to \infty$ and/or $\vert W \vert \to
\infty$. It is shown in Figure 2, upto a constant shift of
different branches in the $y$-direction. 
\begin{figure}[htb]
\begin{center}
\epsfig{file=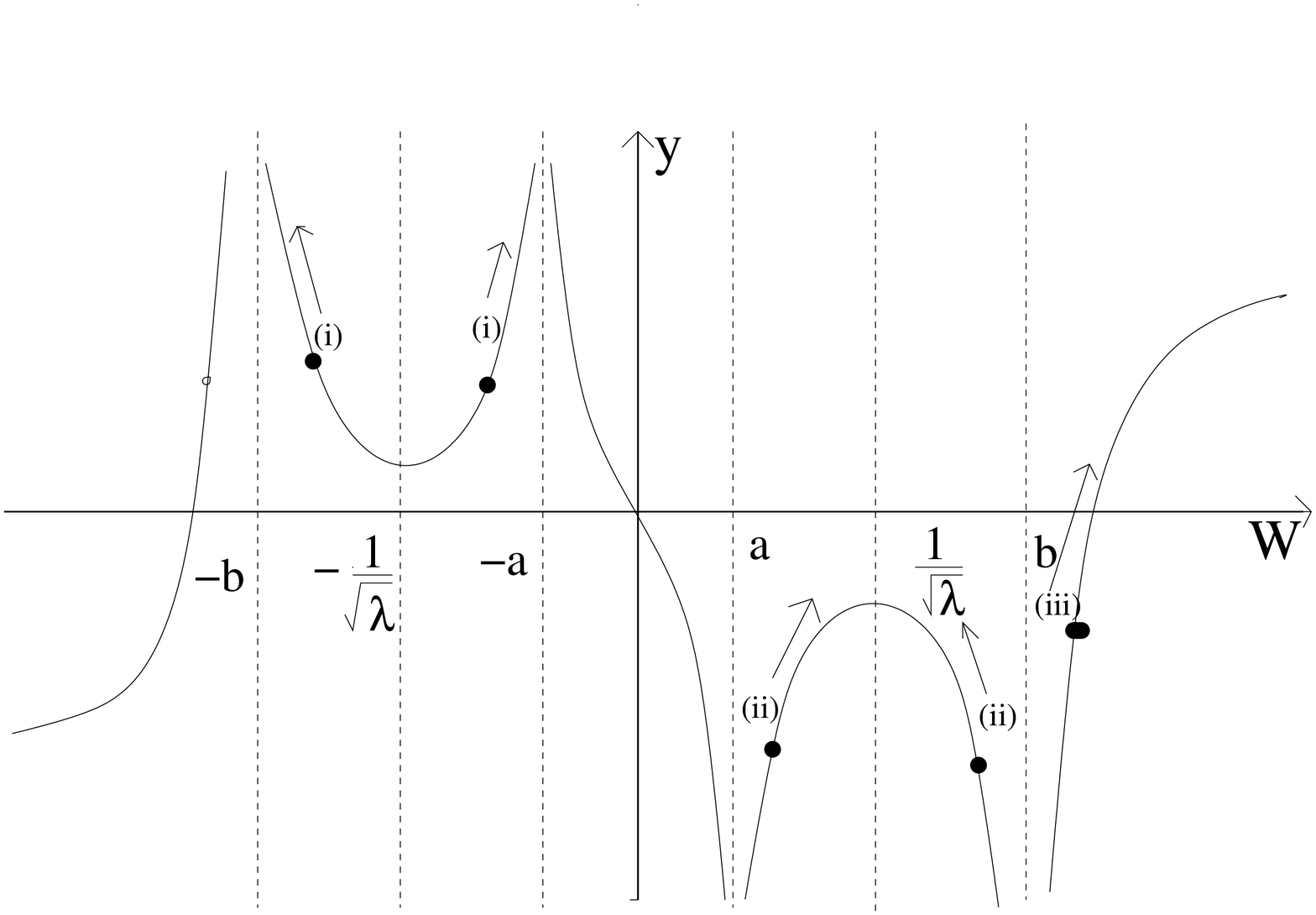, width=12cm, angle=0}
\end{center}
\caption{Plot of $y$ vs $W$, obtained qualitatively from Figure
1. Allowed regions are $|a| \le |W| \le |b|$.} 
\end{figure}

Let $W = W_+$ at $y = 0_+$. Clearly, $W_+$ must lie in the
allowed region.  Consider the evolution for $y > 0$. (A similar
analysis can be done for $y < 0$ with $W= W_-$ at $y = 0_-$.)
Since $y$ must increase, the direction of evolution must be
along the branch which contains $W_+$, and must be in the upward
direction. It is clear, from Figure 2, that either \\ 
{\bf (i)} $y \to \infty$ and $W$ tends to a negative
value, or \\
{\bf (ii)} $y$ tends to a finite value and 
$\frac{d y}{d W} \to 0$, or \\
{\bf (iii)} $y$ tends to a finite value and $W \to \infty$. \\
$W$ itself may increase or decrease depending on its initial
value $W_+$.

In case {\bf (i)}, $M_4 \to \infty$ find W tends to a negative value. See
eq (9). In case {\bf (ii)}, there is a singularity since $W'$ and, hence,
the Ricci scalar $R$ diverge. In case {\bf (iii)} also \footnote{This case
will not arise in the present example since the corresponding branch lies
outside the allowed region. Generically, however, it will.},
there is a singularity since $W$ and, hence, $R$ diverge. Thus,
for $V = - \alpha^2$, it is immediately clear that, irrespective
of the values of $a$ and $b$ in (\ref{bc}) which only determine
$W_\pm$, the solutions will either have a divergent $M_4$, or a
singularity at a finite distance from the brane, or both.

Similar analysis can be performed in all cases where one starts
with a $W(\phi)$. Note that no explicit solutions $\phi(y)$ and
$A(y)$ are needed to determine the presence or absence of
singularities, and the finiteness or otherwise of $M_4$.
Perhaps, the only difficult step is in inverting the given
function $W(\phi)$ to obtain $\phi(W)$, and thus $W_{(1)}(W)$
and $V(W)$, as functions of $W$.

Using this method of analysis, we can easily construct a whole
class of potentials $V(\phi)$ for which there are solutions with
finite $M_4$ and no singularities. One first determines a
suitable function $y(W)$ with the desireable properties:
Clearly, in the $y - W$ plane, there must be one branch where $y
\to \infty$ and $W \to W_R > 0$ with $0 < W_R < \infty$; and
another branch where $y \to - \infty$ and $W \to W_L < 0$ with
$- \infty < W_L < 0$, neither of the branches containing any
critical point where $\frac{d y}{d W} \to 0$. See Figure 3. This
will then ensure that for some continuous, non trivial, ranges
of the parameters $a$ and $b$ in (\ref{bc}), there exist
singularity free solutions with finite $M_4$.
\begin{figure}[htb]
\begin{center}
\epsfig{file=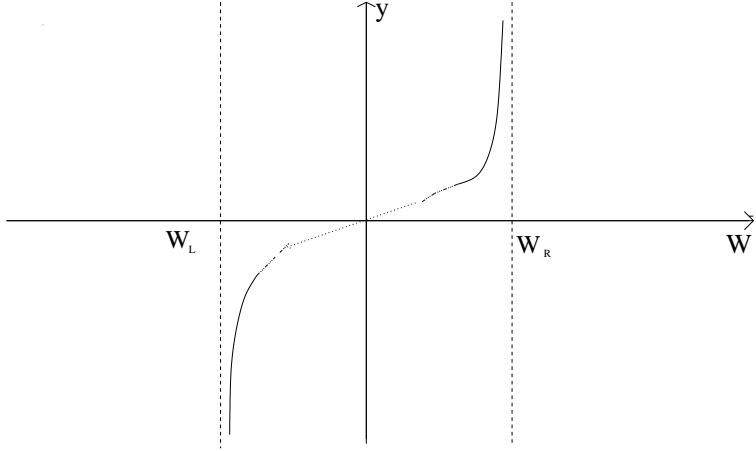, width=10cm, angle=0}
\end{center}
\caption{An example of a function $y(W)$ with desireable
properties.}
\end{figure}

Once a suitable function $y(W)$ is chosen, equations (\ref{w'v})
and (\ref{vw}) will give $V(W)$ and $W_{(1)}(W)$. $V(\phi)$ can
then be obtained by a series of straightforward operations such
as differentiation, integration, functional inversion, etc. 

We now illustrate such a construction by an example. A simple
function $y(W)$, with all the properties mentioned above, is the
one in Figure 3 where the two branches are connected with no
critical point in between, and $W_R = - W_L = \beta$ with $0 <
\beta < \infty$. A simple choice for $y(W)$ is given, for
example, by 
\begin{equation}\label{ywsimple}
\frac{d y}{d W} = \frac{1}{\alpha^2 (\beta^2 - W^2)} \; , 
\end{equation}
with the allowed region given by 
$(\beta^2 - W^2) (1 - \lambda W^2) \ge 0$. 

Let $\lambda = 0$ and $\alpha = \beta = 1$ in equation
(\ref{ywsimple}). It then follows that $V = 1 - 2 W^2$ and
$W_{(1)}^2 = 1 - W^2$. Solving this, one obtains $W = \epsilon
sin \phi$ and $V(\phi) = cos 2 \phi$, where $\epsilon = \pm
1$. The explicit solutions for $\phi(y)$ and $W(y)$ is given by 
\begin{eqnarray}
W & = & \epsilon_\pm sin \phi = 
tanh(y + y_\pm) \nonumber \\
\phi' & = & \delta_\pm sech(y + y_\pm) \label{cos} 
\end{eqnarray}
where $\pm$ signs indicate that the solutions are valid for
positive or negative values of $y$, and the $\epsilon$'s and
$\delta$'s take values $\pm 1$. It is easy to see that these
solutions are singularity free with finite $M_4$. Upon imposing
the boundary conditions (\ref{bc}), $\epsilon_\pm$ and
$\delta_\pm$ can be determined, and $y_\pm$ can be obtained in
terms of $a$ and $b$. However, it turns out that either $a b =
0$ (that is, either $a$, or $b$, or both vanish) or $a^2 + b^2 =
1$. Any of these conditions amounts to a fine tuning. If this
fine tuning relation is not satisfied then, either $M_4$ will
diverge, or there will be singularities, or both. To show this
requires a further analysis which, however, is beyond the scope
of the present work and will be described elsewhere.

Furthermore, solutions in (\ref{cos}) also obey the equation
$W^2 + \phi'^2 = 1$ everywhere, which is not part of the
equations of motion, but is precisely the one obtained from
$V = 1 - 2 W^2$. Note also that the values of $\phi$ at 
$y = \pm \infty$ is given by $\phi_c = n \pi + \frac{\pi}{2}$
and that $V_{(2)}(\phi_c) = - cos(2 \phi_c) > 0$, as required by
the general analysis in section {\bf 3}.

Let $\lambda > 0$, $\alpha^2 = \lambda s^2$, and $\beta^2 =
\lambda^{- 1}$ in equation (\ref{ywsimple}).  It then follows
that $W = s \phi$, which is precisely the case analysed in
\cite{lowzee} where it is shown that there exist singularity free
solutions with finite $M_4$. However, the constants $a$ and $b$
are not arbitrary, but must be fine tuned.  Furthermore, these
solutions also obey the equation $\phi' = s (1 - \lambda W^2)$,
which is not part of the equations of motion but is precisely
the one obtained from equation (\ref{phi'w}) using $W = s \phi$.

Thus, we have a method of constructing the potentials $V(\phi)$,
which admit singularity free solutions with finite $M_4$ for
some continuous, non trivial, ranges of the parameters $a$ and
$b$ in (\ref{bc}). However, there will always be an extra
relation imposed on the arbitrary constants $a$ and $b$. This,
indeed, is fine tuning.

The origin of this extra condition and, thus, of the fine tuning
is clear in the present approach. It arises because $V$ and,
hence, $W_{(1)}$ are specific functions, which depend on the
choice of $y(W)$, which in turn was constructed to ensure that
there exist singularity free solutions with finite $M_4$ for
some continuous, non trivial, ranges of the parameters $a$ and
$b$.  Since $W_{(1)}$ is related to $\phi'$, as given by
(\ref{phi'w}), $W_{(1)} = W_{(1)}(W)$ implies a new equation for
$\phi'$ and $A' \; (= W)$, which is not part of the equations of
motion (\ref{w'}) - (\ref{phi''}) obtained from the action.
\footnote{As noted in the above examples, one gets $\phi'^2 +
W^2 = 1$ in the $\lambda = 0$ case, and $\phi' = s (1 - \lambda
W^2)$ in the $\lambda > 0$ case.} namely $\phi' = (1 - \lambda
{A^\prime}^2) W_{1}(A') $. This is the origin of fine
tuning.

%
%

\vspace{4ex}

{\bf 5.}  
In this paper, we considered an action containing the higher
derivative terms for graviton in the specific Gauss Bonnet
combination and studied whether a singularity free solution with
finite $M_4$ is possible without a fine tuning. We proved that
such a solution is not possible. This is the analogue of the no
go theorem of \cite{csaki}, but now valid for the case where the
action contains the higher derivative Gauss Bonnet terms.

We provided a new method of analysis of the equations
involved in which the qualitative features of the solutions,
such as the presence or absence of singularities, finiteness or
otherwise of $M_4$, etc, can be seen easily without obtaining
the solutions explicitly. This method is applicable quite
generally and provides a constructive way of obtaining
potentials $V(\phi)$ which will admit singularity free solutions
with finite $M_4$. However, such solutions will all require a
fine tuning, consistent with the present no go theorem and that
of \cite{csaki}. The origin of the fine tuning is transparent
in this method.

This no go theorem is likely to be valid for all higher
derivative terms. However, it is not clear how to extend the
proof for the most general case. One hurdle, among possibly many
others, is that the equations of motion will involve derivatives
higher than two. (The case of Gauss Bonnet terms is an
exception.  See \cite{lowzee}.) Also, such terms will invovle
ghosts upon quantisation.  Their physical implication is then
not clear. Nevertheless, considering the importance of the
issue, it is desireable to establish, if true, the present no go
theorem for the most general case. Its failure, if happens, will
also be very interesting as it may provide further insights into
the cosmological constant problem.

The equations of motion analysed here appear in other contexts
also with minor modifications. For example, they appear in the
renormalisation group flow of field theories in the AdS/CFT
correspondence \cite{rgflow} and in the cosmological evolution
of a $(3 + 1)-$dimensional universe \cite{felder}. It will be
of interest to explore these connections and, in particular, to
study the implications of the present no go theorem in the above
mentioned contexts.


\begin{thebibliography}{999}

\bibitem{rs}
L. Randall and R. Sundrum, 
Phys. Rev. Lett. {\bf 83} (1999) 4690, 
hep-th/9906064. 

\bibitem{sundrumkachru}
N. Arkani-Hamed et al, 
Phys. Lett. {\bf B 480} (2000) 193, 
hep-th/0001197; 
S. Kachru, M. Schulz. and E. Silverstein, 
Phys. Rev. {\bf D 62} (2000) 045021, 
hep-th/0001206; 
Phys. Rev. {\bf D 62} (2000) 085003, 
hep-th/0002121.  

\bibitem{weinberg}
See S. Weinberg, 
Rev. Mod. Phys. {\bf 61} (1989) 1 
and the references therein. 

\bibitem{others}
S. Gubser, 
hep-th/0002160. 
S. Forste et al,
Phys. Lett. {\bf B 481} (2000) 360,
hep-th/0002164. 
S. P. de Alwis, 
Nucl. Phys. {\bf B 597} (2001) 263, 
hep-th/0002174; 
C. Gomez, B. Janssen, and P. J. Silva, 
JHEP {\bf 0004} (2000) 027, 
hep-th/0003002. 
S. P. de Alwis, A. T. Flournoy, and N. Irges, 
JHEP {\bf 0101} (2001) 027, 
hep-th/0004125. 
G. T. Horowitz, I. Low, and A. Zee, 
Phys. Rev. {\bf D 62} (2000) 086005, 
hep-th/0004206. 

\bibitem{csaki}
C. Csaki et al, 
Nucl. Phys. {\bf B 584} (2000) 359, 
hep-th/0004133. 

\bibitem{lowzee}
I. Low and A. Zee, 
Nucl. Phys. {\bf B 585} (2000) 395, 
hep-th/0004124. 

\bibitem{neup}
I. P. Neupane,
hep-th/0106100 (To appear in JHEP)

\bibitem{gbothers}   
I. P. Neupane,
hep-th/0008190 (Published in JHEP 09 (2000)040);
Y.M. Cho, I.P. Neupane and P.S. Wesson,
hep-th/0104226 (Published in Nucl. Phys. B, 621(2002) 388-412;
Cristiano Germani, Carlos F. Sopuerta,
hep-th/0202060,Phys. Rev. Lett. 88, 231101 (2002);
Christos Charmousis, Jean-Francois Dufaux,
hep-th/0202107;

\bibitem{zumino}
B. Zwiebach, 
Phys. Lett. {\bf B 156} (1985) 315; 
B. Zumino, 
Phys. Rep. {\bf 137} (1986) 109. 

\bibitem{gubser}
O. DeWolfe et al, 
Phys. Rev. {\bf D 62} (2000) 046008, 
hep-th/9909134. 
See also K. Skenderis and P. K. Townsend, 
Phys. Lett. {\bf B 468} (1999) 46,hep-th/9909070 

\bibitem{rgflow} 
See for example, E.Verlinde and H.Verlinde
JHEP 005(2000)034, hep-th/9912018 .
See also the recent review 
E. D'Hoker and D. Z. Freedman, 
hep-th/0201253 and the references therein. 

\bibitem{felder} 
G. Felder, A. Frolov, and L. Kofman, 
hep-th/0112165; 
G. Felder, A. Frolov, L. Kofman, and A. Linde, 
hep-th/0202017. 

\end{thebibliography}
\end{document}